\documentclass[superscriptaddress,twocolumn,showpacs,prb]{revtex4-2}
\usepackage[utf8]{inputenc}
\usepackage{amsmath}
\usepackage{mathtools}
\usepackage{braket}
\usepackage{graphicx}
\usepackage{amsfonts}
\usepackage{pgfplots}
\usepackage{csquotes}
\usepackage{hhline}
\usepackage{amssymb}
\usepackage{listings}
\usepackage{color}
\usepackage{xcolor}

\usepackage{tikz}
\usetikzlibrary{quantikz}

\makeatletter
\newcommand{\ssymbol}[1]{^{\@fnsymbol{#1}}}
\makeatother

\usepackage{lipsum}
\definecolor{codegreen}{rgb}{0,0.6,0}
\definecolor{codegray}{rgb}{0.5,0.5,0.5}
\definecolor{codepurple}{rgb}{0.58,0,0.82}
\definecolor{backcolour}{rgb}{0.95,0.95,0.92}
 
\lstdefinestyle{mystyle}{
    backgroundcolor=\color{backcolour},   
    commentstyle=\color{codegreen},
    keywordstyle=\color{magenta},
    numberstyle=\tiny\color{codegray},
    stringstyle=\color{codepurple},
    basicstyle=\footnotesize,
    breakatwhitespace=false,         
    breaklines=true,                 
    captionpos=b,                    
    keepspaces=true,                 
    numbers=left,                    
    numbersep=5pt,                  
    showspaces=false,                
    showstringspaces=false,
    showtabs=false,                  
    tabsize=2
}
 
\usepackage{subfigure}

\usepackage{dcolumn}
\usepackage{tabularx}
\usepackage[colorlinks=true,linkcolor=blue,citecolor=blue,urlcolor=blue]{hyperref}
\usepackage{longtable}
\usepackage{braket}
\usepackage{float}
\newcolumntype{C}{>{\centering\arraybackslash}X}
\usepackage{multirow}

\usepackage{tikz,xcolor,hyperref}
\definecolor{lime}{HTML}{A6CE39}
\DeclareRobustCommand{\orcidicon}{%
	\begin{tikzpicture}
	\draw[lime, fill=lime] (0,0) 
	circle [radius=0.16] 
	node[white] {{\fontfamily{qag}\selectfont \tiny ID}};
	\draw[white, fill=white] (-0.0625,0.095) 
	circle [radius=0.007];
	\end{tikzpicture}
	\hspace{-2mm}}
\foreach \x in {A, ..., Z}{%
	\expandafter\xdef\csname orcid\x\endcsname{\noexpand\href{https://orcid.org/\csname orcidauthor\x\endcsname}{\noexpand\orcidicon}}}

\pgfplotsset{compat=1.17}

\begin{document}
\title{A secure deterministic remote state preparation via a seven-qubit entangled channel of an arbitrary two-qubit state under the impact of quantum noise}

\author{Deepak Singh \orcidA{}}
\email{dsingh@ma.iitr.ac.in}
\affiliation{Department of Mathematics,\\ Indian Institute of Technology Roorkee 247667, Uttarakhand, India}

\author{Sanjeev Kumar \orcidB{}}
\email{sanjeev.kumar@ma.iitr.ac.in}
\affiliation{Department of Mathematics, Mehta Family School of Data Science and Artificial Intelligence\\ Indian Institute of Technology Roorkee 247667, Uttarakhand, India}

\author{Bikash K. Behera \orcidC{}}
\email{bikas.riki@gmail.com}
\affiliation{Bikash's Quantum (OPC) Pvt. Ltd., Balindi, Mohanpur 741246, Nadia, West Bengal, India}
                                                                        
\begin{abstract}
As one of the most prominent subfields of quantum communication research, remote state preparation (RSP) plays a crucial role in quantum networks. Here we present a deterministic remote state preparation scheme to prepare an arbitrary two-qubit state via a seven-qubit entangled channel created from Borras \emph{et al.} state. Quantum noises are inherent to each and every protocol for quantum communication that is currently in use, putting the integrity of quantum communication systems and their dependability at risk. The initial state of the system was a pure quantum state, but as soon as there was any noise injected into the system, it transitioned into a mixed state. In this article, we discuss the six different types of noise models namely bit-flip noise, phase-flip noise, bit-phase-flip noise, amplitude damping, phase damping and depolarizing noise. The impact these noises had on the entangled channel may be seen by analysing the density matrices that have been altered as a result of the noise. For the purpose of analysing the impact of noise on the scheme, the fidelity between the original quantum state and the remotely prepared state has been assessed and graphically represented. In addition, a comprehensive security analysis is performed, demonstrating that the suggested protocol is safe against internal and external attacks.
\end{abstract}


\maketitle
\onecolumngrid
\section{Introduction}\label{sec:introduction}\label{Section-1}
Quantum communication is a very essential field in quantum mechanics, quantum computation, and quantum information theory. Entanglement is the backbone of quantum communication; it is considered the essential asset of overall quantum information processing \cite{nielsen2001quantum}. 
It is well known that there are some novel phenomena in the applications of quantum communication and information theory, which include quantum teleportation \cite{bennett1993teleporting, boschi1998experimental}, quantum secure direct communication (QSDC)\cite{long2007quantum}, quantum key distribution \cite{scarani2009security} and quantum dense coding \cite{mattle1996dense}, quantum secret sharing \cite{gisin2002quantum, zhang2005multiparty}, quantum data hiding \cite{verstraete2003quantum}, quantum private comparison (QPC) \cite{yang2013corrigendum}, quantum remote state preparation (RSP) \cite{bennett2001remote, bennett2005remote, zhang2016deterministic}  to name a few. 
Using an entangled channel to send the quantum state securely in all of these quantum communication protocols is an important area of research in quantum communication. At least in theory, quantum protocols have the potential to attain a better degree of security than their conventional counterparts.

In a remote state preparation (RSP) protocol, the sender uses a shared entangled channel and the right measurements to set up a known quantum state for the faraway receiver. RSP is believed to be a more efficient method of teleporting to a known state than the traditional method since it uses fewer classical bits \cite{pati2000minimum}. Since then, several distinct RSP algorithms \cite{berry2003optimal, devetak2001low} were put up as potential solutions. Later on, a number of research that were based on RSP were carried out. These included joint remote state preparation (JRSP) \cite{chen2011joint, hou2009joint, yang2012joint} and controlled remote state preparation (CRSP) [\cite{chen2012controlled}. On the other hand, the vast majority of RSP algorithms made in the past have a chance of success that is less than 1. To raise the success probability of RSP to 1, a new RSP algorithm called the deterministic remote state preparation (DRSP) algorithm \cite{zhang2017deterministic, ma2017deterministic} was proposed. This algorithm can set up the needed quantum state with a one-in-one chance of success, which saves a lot of quantum resources. In recent years, several fascinating studies about RSP have been carried out. A unique technique for the implementation of RSP of a generic m-qubit entangled state was suggested by Wang et al. \cite{wang2015generalized} by employing the GHZ-type states as quantum channels with a high success probability, reducibility, and generalizability. This strategy was developed by using the GHZ-type states. Wang et al.\cite{wang2015efficient}  suggested two successful measurement-based ways for executing the RSP techniques for generic W-class entangled states for three and four particles, employing GHZ-type states as the quantum channels in the same year.
In comparison to the previous schemes, these approaches stand a better chance of being implemented successfully with a higher success probability compared to the existing schemes. In addition, the suggested methods may be successfully implemented with a total success probability of one when the utilised channels are condensed into the most maximally entangled versions of themselves. Wang et al. \cite{wang2016practical} proposed two optics-based implementations for RSP and JRSP of an arbitrary single-photon pure state. The protocols for these implementations may be achieved with a specific success probability with the assistance of appropriate LOCC.

Every quantum protocol is practically not error-free. It has some ambiguities in the system. These ambiguities are considered quantum noise. Quantum noise is inevitable in implementing a quantum communication system under realistic conditions. Quantum noise will hurt the safety and reliability of the quantum communication system in a big way. Research on RSP \cite{wang2017effect, liang2015effects} in noisy environments is now being conducted and explored. The JRSP method was computed in amplitude-damping noise, and phase-damping noise by Guan et al. \cite{guan2014joint}, who also conducted an in-depth analysis of the impact that noise has on the output state. Ma et al. \cite{ma2017deterministic} examined the influence of amplitude-damping and phase-damping noise on the DJRSP method and presented a deterministic approach for constructing distant states using a Brown et al. state as the underlying channel. In \cite{dash2020deterministic} the noisy environment is studied for deterministic joint remote state preparation of arbitrary two-qubit state. A complexity analysis of the teleportation scheme under the influence of noise is studied in \cite{singh2021complexity}. 

Motivated by these schemes, we have developed a protocol for the deterministic remote state preparation of a two-qubit qubit quantum state via a maximally entangled seven-qubit state that is derived from Borras \emph{et al.} state. The use of highly entangled Borras \emph{et al.} \cite{borras2007multiqubit} state in the remote state preparation has not been done in any of the schemes proposed till now.  This scheme involves four participants in preparation for the remote state. Alice, Bob, Charlie, and David share a predetermined seven-qubit entangled channel. Alice has the first qubit, Bob has the second and third, Charlie has the fourth and sixth, and David has the last two qubits, which are the fifth and seventh. The first action that has to be performed is to factorize the entangled channel into the sum of the Bell basis states that have the same coefficients as the two-qubit state that needs to be prepared remotely. Then Alice measures her qubit and conveys the result to Bob through a classical channel. Charlie and David likewise communicate their outcomes to Bob via classical channels. Finally, in order to remotely prepare the desired two qubit-state, Bob needs to perform certain unitary operation on his qubit based on the collapse state of other participants. All the recovery operations are discussed in details in table \ref{table:Operation}. Next, we will study the impact of six types of noise on the entangled channel. These six kinds of noise are named bit-flip noise, phase-flip noise, bit-phase-flip noise, amplitude damping, phase damping and depolarizing noise. These noise models are studied with the help of the action of the Kraus operator on the qubit. When the Kraus operator acts on a quantum state, it becomes a mixed state. Upon evaluation of the density matrices, the fidelity is calculated between the initial state and the remotely prepared state. The variation in fidelity is  represented with the help of a graph. In \cite{chen2012controlled}, the security analysis of the remote state preparation protocol is performed. This study analyses the security attack from the outside and the inside participants. 

The rest of the article is structured as follows: Section \ref{Section-2} covers introductory terminology about the underlying entangled channel and remote state preparation procedures. In Section \ref{Section-3}, the noise analysis of the entangled teleportation channel is described. Section \ref{Section-5} consists of the security analysis of the protocol, and \ref{Section-5} concludes the report with a discussion of the study's results, followed by suggestions for further research.

\section{Remote state preparation of an arbitrary two-qubit state}\label{Section-2}
Suppose Alice, Bob, Charlie, and David are the participants in this system. Alice, Charlie and David combining together, want to remotely prepare a known quantum state at Bob's end. The two-qubit quantum state can be parameterized as $\ket{\xi} = \alpha \ket{00} + \beta \ket{11}$,  where $ \alpha,\beta \in \mathbb{C}$ such that $|\alpha|+|\beta|^2 = 1$ takes care of the normalization of the state $\Ket{\phi}$.
The seven-qubit entangled channel shared among the participants can be prepared by using the Borras \emph{et al.} state and an ancilla qubit in the $\Ket{0}$ state. The Borras \emph{et al.} state is given by 
\begin{eqnarray}
    \ket{\psi} &=& \frac{1}{4}\Big(\ket{000}(\ket{0}\ket{\psi^{+}} + \ket{1}\ket{\phi^{+}}) +\ket{001}(\ket{0}\ket{\phi^{-}} - \ket{1}\ket{\psi^{-}})
    +\ket{010}(\ket{0}\ket{\phi^{+}} - \ket{1}\ket{\psi^{+}}) + \ket{011}(\ket{0}\ket{\psi^{-}} + \ket{1}\ket{\phi^{-}}) \nonumber\\
    &-& \ket{100}(\ket{0}\ket{\phi^{-}} + \ket{1}\ket{\psi^{-}}) +\ket{101}(-\ket{0}\ket{\psi^{+}} + \ket{1}\ket{\phi^{+}}) + \ket{110}(\ket{0}\ket{\psi^{-}} - \ket{1}\ket{\phi^{-}}) + \ket{111}(\ket{0}\ket{\phi^{+}} + \ket{1}\ket{\psi^{+}}) \Big) \nonumber \\
\label{eq-Borras}    
\end{eqnarray}

where $\ket{\phi^{\pm}} = \frac{1}{\sqrt{2}} (\ket{01} \pm \ket{10})$ and $\ket{\psi^{\pm}} = \frac{1}{\sqrt{2}} (\ket{00} \pm \ket{11})$. 
The ancilla qubit is entangled with the Borras \emph{et al.} state and the final entangled channel is given by the following equation
\begin{eqnarray}
    \ket{\Psi} &=& \Big( \ket{\psi}_{123456} \otimes \ket{0}_{7}  \Big)_{CX(6,7)} \nonumber \\
    &=& \frac{1}{4\sqrt{2}}\Big( \ket{0000000}
+\ket{0000111}
+\ket{0001011}
+\ket{0001100}
+\ket{0010011}
-\ket{0010100}
-\ket{0011000}
+\ket{0011111} \nonumber \\
&+& \ket{0100011}
+\ket{0100100}
-\ket{0101000}
-\ket{0101111}
+\ket{0110000}
-\ket{0110111}
+\ket{0111011}
-\ket{0111100}
-\ket{1000011} \nonumber \\
&+& \ket{1000100}
-\ket{1001000}
+\ket{1001111}
-\ket{1010000}
-\ket{1010111}
+\ket{1011011}
+\ket{1011100}
+\ket{1100000}
-\ket{1100111} \nonumber \\
&-& \ket{1101011}
+\ket{1101100}
+\ket{1110011}
+\ket{1110100}
+\ket{1111000}
+\ket{1111111}
 \Big)_{AB_1B_2C_1D_1C_2D_2}
\label{Borras_Eq1}    
\end{eqnarray}
The Borras \emph{et al.} state is a  six-qubit highly entangled state, which included two-qubit maximally entangled Bell states, $\ket{\phi^{\pm}}$ and $\ket{\psi^{\pm}}$ \cite{borras2007multiqubit}. Because the prepared entangled state $\ket{\Psi}$ was generated using the C-NOT operation on a highly entangled six-qubit Borras et al. state, which is considered to add entanglement to a quantum state, we believe that it is indeed extremely  entangled, if not maximally entangled. Once the entangled channel is obtained, Alice reserves the first qubit for herself and distributes the other qubits as follows: Bob receives the second and third qubits, Charlie receives the fourth and sixth qubits, and David receives the fifth and seventh qubits. Alice owns the qubit $A$, Bob owns the qubits $B_1$ and $B_2$, Charlie owns the qubits $C_1$ and $C_2$, and David owns the qubits $D_1$ and $D_2$.
In order to remotely prepare the quantum state at Bob's end, the factorization of the quantum state plays a crucial role.
So, the known state $\ket{\Psi}$ is primarily factorized in a certain way that only the sender utilises a basis created out of the known parameters $\alpha$ and $\beta$. Here the factorization of $\ket{\Psi}$ can be done in the following basis -

\begin{figure}[ht]
     \centering
     \subfigure[]{\includegraphics[width=0.45\textwidth]{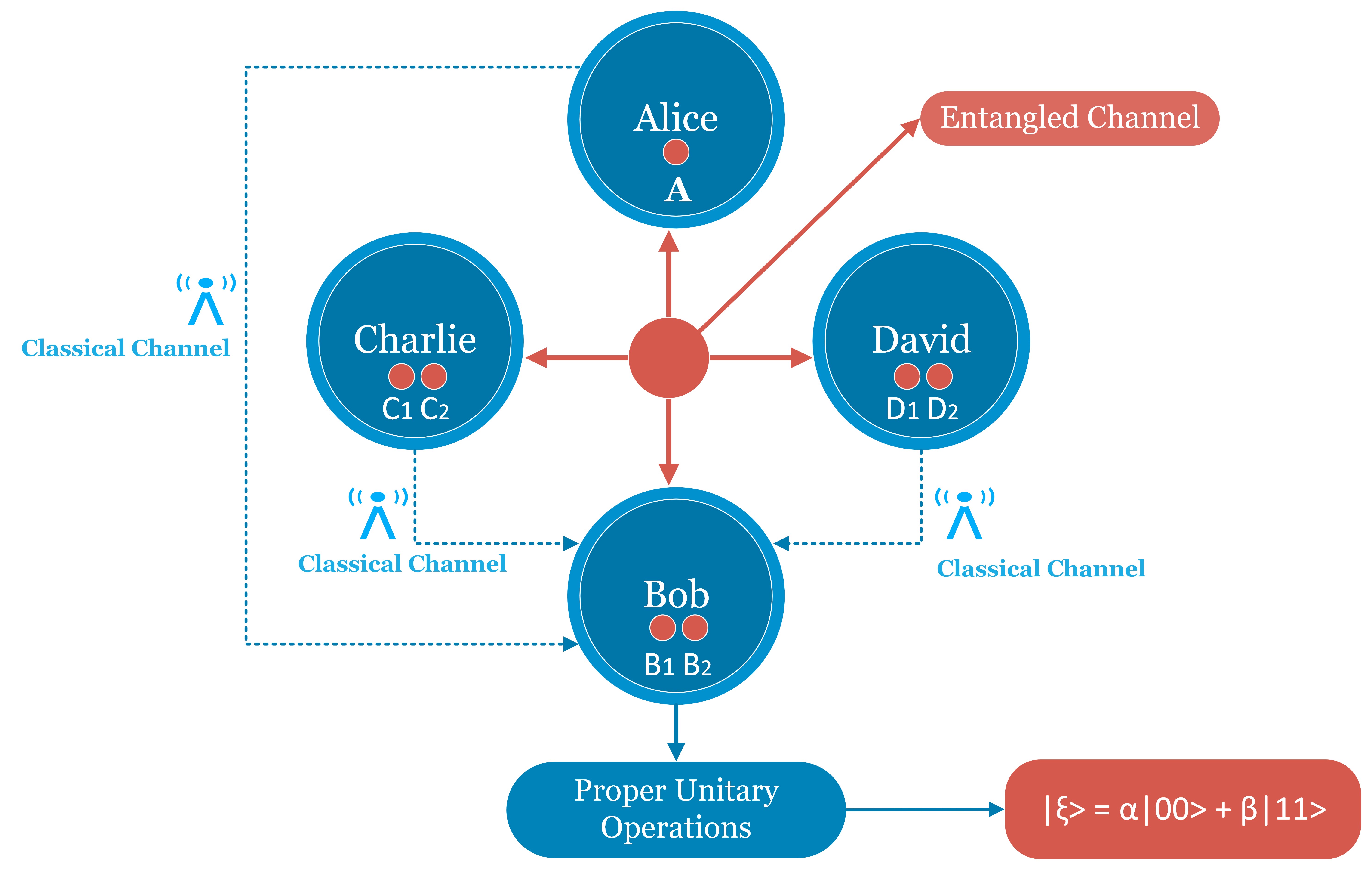}} \label{fig:Graph Plot 1}
     \subfigure[]{\includegraphics[width=0.45\textwidth]{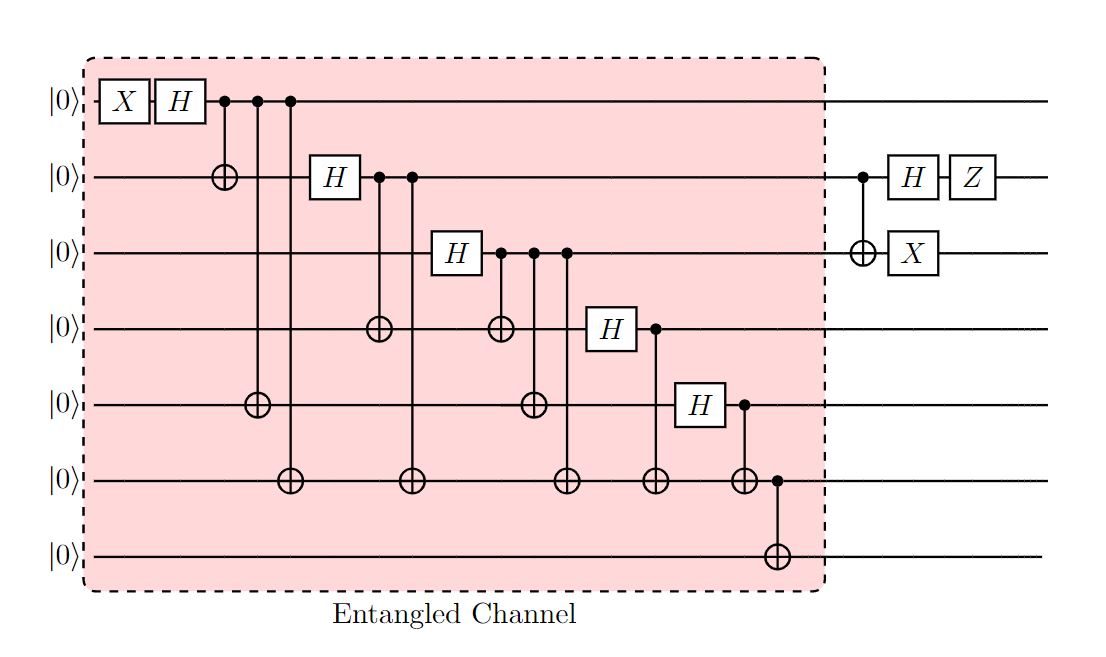}} \label{fig:Graph Plot 2}
     \caption{ (a) Diagram representation of the remote state preparation protocol (b) Quantum Circuit representation of the RSP scheme. The } 
     \label{fig:Figure}
 \end{figure}

\begin{eqnarray}
    \ket{\Psi} &=& \dfrac{1}{4} \Big[ \ket{\Upsilon_1}\ket{\zeta_1} + \ket{\Upsilon_2}\ket{\zeta_2}  \Big]_{AB_1B_2B_2C_1C_2D_1D_2}
\label{Eq2}    
\end{eqnarray}
where the quantum state $\ket{\Upsilon_1}$ and $\ket{\Upsilon_1}$ are given by the Eq.\eqref{Eq3} 
\begin{eqnarray}
    \ket{\Upsilon_1}_A &=& (\alpha \Ket{0} + \beta \Ket{1})_{A} \nonumber \\ 
    \ket{\Upsilon_2}_A &=& (\alpha \Ket{1} - \beta \Ket{0})_{A}
\label{Eq3}    
\end{eqnarray}

And the quantum states $\Ket{\zeta_1}$ and $\Ket{\zeta_2}$ are given by the following expressions

\begin{eqnarray}
    \ket{\zeta_1} = \dfrac{1}{\sqrt{2}} &\Big[& \big( \alpha \Ket{00} + \beta \Ket{10} + \alpha \Ket{11}-\beta \Ket{01} \big)_{B_1B_2} \Ket{00}_{C_1C_2}\Ket{00}_{D_1D_2} \nonumber\\
    &+& \big(\alpha \Ket{01} + \beta \Ket{11} + \alpha \Ket{10}-\beta \Ket{00}\big)_{B_1B_2} \Ket{01}_{C_1C_2} \Ket{01}_{D_1D_2} \nonumber \\
    &+& \big(-\alpha \Ket{01} + \beta \Ket{11} - \alpha \Ket{10}-\beta \Ket{00}\big)_{B_1B_2} \Ket{10}_{C_1C_2} \Ket{00}_{D_1D_2} \nonumber \\
    &+& \big(\alpha \Ket{00} - \beta \Ket{10} + \alpha \Ket{11}+\beta \Ket{01}\big)_{B_1B_2} \Ket{11}_{C_1C_2} \Ket{01}_{D_1D_2} \nonumber \\ 
    &+& \big(-\alpha \Ket{01} + \beta \Ket{11} + \alpha \Ket{10}+\beta \Ket{00}\big)_{B_1B_2} \Ket{00}_{C_1C_2} \Ket{10}_{D_1D_2} \nonumber \\
    &+& \big(\alpha \Ket{00} + \beta \Ket{10} - \alpha \Ket{11} + \beta \Ket{01}\big)_{B_1B_2} \Ket{10}_{C_1C_2} \Ket{10}_{D_1D_2} \nonumber \\
    &+& \big(\alpha \Ket{00} - \beta \Ket{10} - \alpha \Ket{11} - \beta \Ket{01}\big)_{B_1B_2} \Ket{01}_{C_1C_2} \Ket{11}_{D_1D_2} \nonumber \\
    &+& \big(\alpha \Ket{01} + \beta \Ket{11} - \alpha \Ket{10} + \beta \Ket{00}\big)_{B_1B_2} \Ket{11}_{C_1C_2} \Ket{11}_{D_1D_2} \Big]
    \label{Eq01}    
\end{eqnarray}
\begin{eqnarray}
    \ket{\zeta_2} = \dfrac{1}{\sqrt{2}} &\Big[& \big(\alpha \Ket{10} - \beta \Ket{00} - \alpha \Ket{01}-\beta \Ket{11}\big)_{B_1B_2} \Ket{00}_{C_1C_2}\Ket{00}_{D_1D_2} \nonumber\\
    &+& \big(\alpha \Ket{11} - \beta \Ket{01} - \alpha \Ket{00}-\beta \Ket{10}\big)_{B_1B_2} \Ket{01}_{C_1C_2} \Ket{01}_{D_1D_2} \nonumber \\
    &+& \big(\alpha \Ket{11} + \beta \Ket{01} - \alpha \Ket{00}+\beta \Ket{10}\big)_{B_1B_2} \Ket{10}_{C_1C_2} \Ket{00}_{D_1D_2} \nonumber \\
    &+& \big(-\alpha \Ket{10} - \beta \Ket{00} + \alpha \Ket{01}-\beta \Ket{11}\big)_{B_1B_2} \Ket{11}_{C_1C_2} \Ket{01}_{D_1D_2} \nonumber \\ 
    &+& \big(\alpha \Ket{11} + \beta \Ket{01} + \alpha \Ket{00} - \beta \Ket{10}\big)_{B_1B_2} \Ket{00}_{C_1C_2} \Ket{10}_{D_1D_2} \nonumber \\
    &+& \big(\alpha \Ket{10} - \beta \Ket{00} + \alpha \Ket{01} + \beta \Ket{11}\big)_{B_1B_2} \Ket{10}_{C_1C_2} \Ket{10}_{D_1D_2} \nonumber \\
    &+& \big(- \alpha \Ket{10} - \beta \Ket{00} - \alpha \Ket{01} + \beta \Ket{11}\big)_{B_1B_2} \Ket{01}_{C_1C_2} \Ket{11}_{D_1D_2} \nonumber \\
    &+& \big(\alpha \Ket{11} - \beta \Ket{01} + \alpha \Ket{00} + \beta \Ket{10}\big)_{B_1B_2} \Ket{11}_{C_1C_2} \Ket{11}_{D_1D_2} \Big]
\label{Eq02}    
\end{eqnarray}
Now Alice measure her qubit in the basis $\{ \ket{\Upsilon_1}, \ket{\Upsilon_1} \}$, Charlie and David measure their qubits in $\{\Ket{00}, \Ket{01}, \Ket{10}, \Ket{11} \}$ basis then Bob can easily remotely prepare the state $\Ket{\xi}$ knowing the outcome of Alice, Charlie and David. For instance, if Alice's measurement collapse in the state $\Ket{\Upsilon_1}_A$, Charlie and David's measurement collapse to the state $\Ket{01}_{C_1C_2}\Ket{01}_{D_1D_2}$ after then, each participant will use a conventional channel to communicate their results to Bob. Bob will next be required to do the necessary unitary operations on his qubit in order to remotely prepare the state at his end., in this case, Bob will apply a controlled-not gate form ``his first qubit" to the second qubit ($B_1$ to $B_2$), denoted as $CX_{1-2}$. A Hadamard gate on his first qubit, denoted as $H(1)$, again a not operation on his second qubit, denoted as $X_2$ and finally a $Z_1$ gate on his first qubit to flip the phase of the first qubit. At this stage of the protocol, Bob has successfully prepared the quantum state $\Ket{\xi}$ at his end. All the other possible cases are given in table \ref{table:Operation}.


\begin{table*}[!htb]
\centering
\renewcommand{\arraystretch}{1.65}
\begin{tabular}{|l|l|l|c|}
\hline
\textbf{Alice's Outcome} & \textbf{Charlie \& David's Outcome} & \textbf{Bob's Collapse State} & \textbf{Bob's Operations} \\ \hline
$\ket{\Upsilon_1}_A$ & $\Ket{00}_{C_1C_2}\Ket{00}_{D_1D_2}$ & $\frac{1}{\sqrt{2}}(\alpha \Ket{00} + \beta \Ket{10} + \alpha \Ket{11} - \beta \Ket{01})$ & $CX_{1-2}~H_1~Z_{1}$ \\ \hline
$\ket{\Upsilon_1}_A$ & $\Ket{01}_{C_1C_2}\Ket{01}_{D_1D_2}$ & $\frac{1}{\sqrt{2}}(\alpha \Ket{01} + \beta \Ket{11} + \alpha \Ket{10}-\beta \Ket{00})$ & $CX_{1-2}~H_1~X_2~Z_{1}$ \\ \hline
$\ket{\Upsilon_1}_A$ & $\Ket{10}_{C_1C_2}\Ket{00}_{D_1D_2}$ & $\frac{1}{\sqrt{2}}(-\alpha \Ket{01} + \beta \Ket{11} - \alpha \Ket{10}-\beta \Ket{00})$ & $CX_{1-2}~H_1~Z_{1}~Z_{2}~X_{2}$ \\ \hline
$\ket{\Upsilon_1}_A$ & $\Ket{11}_{C_1C_2}\Ket{01}_{D_1D_2}$ & $\frac{1}{\sqrt{2}} (\alpha \Ket{00} - \beta \Ket{10} + \alpha \Ket{11}+\beta \Ket{01} )$ & $CX_{1-2}~H_1$ \\ \hline
$\ket{\Upsilon_1}_A$ & $\Ket{00}_{C_1C_2}\Ket{10}_{D_1D_2}$ & $\frac{1}{\sqrt{2}} (-\alpha \Ket{01} + \beta \Ket{11} + \alpha \Ket{10}+\beta \Ket{00} )$ & $CX_{1-2}~H_1~Z_{1}~X_{1}~X_{2}$ \\ \hline
$\ket{\Upsilon_1}_A$ & $\Ket{10}_{C_1C_2}\Ket{10}_{D_1D_2}$ & $\frac{1}{\sqrt{2}} (\alpha \Ket{00} + \beta \Ket{10} - \alpha \Ket{11}+\beta \Ket{01})$ & $CX_{1-2}~H_1~Z_{2}~X_{1}~CX_{2-1}$ \\ \hline
$\ket{\Upsilon_1}_A$ & $\Ket{10}_{C_1C_2}\Ket{11}_{D_1D_2}$ & $\frac{1}{\sqrt{2}} ( \alpha \Ket{00} - \beta \Ket{10} - \alpha \Ket{11} - \beta \Ket{01} )$ & $CX_{1-2}~H_1~Z_{2}~X_{1}$ \\ \hline
$\ket{\Upsilon_1}_A$ & $\Ket{11}_{C_1C_2}\Ket{11}_{D_1D_2}$ & $\frac{1}{\sqrt{2}} (\alpha \Ket{01} + \beta \Ket{11} - \alpha \Ket{10} + \beta \Ket{00} )$ & $CX_{1-2}~H_1~X_{2}~X_{1}$ \\ \hline 
\hline
$\ket{\Upsilon_2}_A$ & $\Ket{00}_{C_1C_2}\Ket{00}_{D_1D_2}$ & $\frac{1}{\sqrt{2}} (\alpha \Ket{10} - \beta \Ket{00} - \alpha \Ket{01} - \beta \Ket{11})$ & $CX_{1-2}~H_1~Z_{1}~X_{1}~X_{2}~Z_{1}$ \\ \hline
$\ket{\Upsilon_2}_A$ & $\Ket{01}_{C_1C_2}\Ket{01}_{D_1D_2}$ & $\frac{1}{\sqrt{2}} (\alpha \Ket{11} - \beta \Ket{01} - \alpha \Ket{00} - \beta \Ket{10} )$ & $CX_{1-2}~H_1~Z_{2}~Z_{1}~X_{1}$ \\ \hline
$\ket{\Upsilon_2}_A$ & $\Ket{10}_{C_1C_2}\Ket{00}_{D_1D_2}$ & $\frac{1}{\sqrt{2}}(\alpha \Ket{11} + \beta \Ket{01} - \alpha \Ket{00} + \beta \Ket{10})$ & $CX_{1-2}~H_1~Z_{1}~X_{1}$ \\ \hline
$\ket{\Upsilon_2}_A$ & $\Ket{11}_{C_1C_2}\Ket{01}_{D_1D_2}$ & $\frac{1}{\sqrt{2}} (-\alpha \Ket{10} - \beta \Ket{00} + \alpha \Ket{01} - \beta \Ket{11} )$ & $CX_{1-2}~H_1~X_{1}~X_{2}~Z_{1}$ \\ \hline
$\ket{\Upsilon_2}_A$ & $\Ket{00}_{C_1C_2}\Ket{10}_{D_1D_2}$ & $\frac{1}{\sqrt{2}} (\alpha \Ket{11} + \beta \Ket{01} + \alpha \Ket{00} - \beta \Ket{10} )$ & $CX_{1-2}~H_1$ \\ \hline
$\ket{\Upsilon_2}_A$ & $\Ket{10}_{C_1C_2}\Ket{10}_{D_1D_2}$ & $\frac{1}{\sqrt{2}} (\alpha \Ket{10} - \beta \Ket{00} + \alpha \Ket{01} + \beta \Ket{11} )$ & $CX_{1-2}~H_1~Z_{1}~X_{2}$ \\ \hline
$\ket{\Upsilon_2}_A$ & $\Ket{10}_{C_1C_2}\Ket{11}_{D_1D_2}$ & $\frac{1}{\sqrt{2}} (-\alpha \Ket{10} - \beta \Ket{00} - \alpha \Ket{01} + \beta \Ket{11} )$ & $CX_{1-2}~H_1~Z_{1}~Z_{2}~X_{2}$ \\ \hline
$\ket{\Upsilon_2}_A$ & $\Ket{11}_{C_1C_2}\Ket{11}_{D_1D_2}$ & $\frac{1}{\sqrt{2}} (\alpha \Ket{11} - \beta \Ket{01} + \alpha \Ket{00} + \beta \Ket{10} )$ & $CX_{1-2}~H_1~Z_{1}$ \\ \hline
\end{tabular}
\caption{Here are the operations that Bob must do on his qubits in order to obtain a state based on the results of Alice and Charlie/David. Here $CX_{1-2}$ represents the CNOT gate from qubit 1 to 2, $H_1$ represents the hadamard gate on qubit 1, $Z_2$ represents the Z gate on qubit 2, and $X_2$ represents the not operation on qubit 2.}
\label{table:Operation}
\end{table*}


\section{Effect of noisy environment on the quantum channel }\label{Section-3}
The influence that noise has on the quantum channel will be the topic of our next discussion. Ambiguities are produced in the quantum system as a result of the fact that a real quantum system does not operate in perfect circumstances and that it interacts with its surrounding environment. This kind of ambiguity is described by the phrase quantum noise. Quantum noise must be taken into account in order to undertake precise research on a quantum communication method.In any actual quantum experiment conducted in the real world, noise will play a crucial role in deciding the success of the scheme. These noises may be studied and classified into one of six categories. Bit flip, phase flip, bit-phase flip, amplitude damping, phase damping, and depolarizing noise are the six fundamental forms of quantum noise that may occur in quantum channels. In this section, these six forms of quantum noise are examined. Therefore, the most productive course of action would be to investigate the causes and consequences of the system's noises and to seek to reduce their impact wherever feasible. We can analyze this by studying the evolution of the density matrix $\rho=\ket{\psi}\bra{\psi}$  with the help of Kraus operators. The impact of noise on the quantum channel can be done using the operator sum representation.  The action of  Kraus operators $E_k$ on a particular qubit $k$ described by the density matrix $\rho_k$ is given by the Eq. \eqref {eq:Xi}
\begin{equation}
    \mathcal{G}^r(\rho_k) = \sum_{j=1}^{n} \big({E_j}\big)\ \rho_k\ \big({E_{j}}\big)^{\dagger}
\label{eq:Xi}
\end{equation}
where $r \in \{b,w,f,a,p,d\}$ for bit-flip, phase-flip, bit-phase-flip, amplitude damping, phase damping and depolarizing damping respectively, $j \in \{0,1\}$ for $r=b,w,f,a$, $j \in \{0,1,2\}$ for $r=p$ and $j \in \{0,1,2,3\}$ for $r=d$.

Alice retains the first qubit and transfers the remaining qubits to Bob, Charlie, and David. After the distribution of qubits in a noisy environment, the shared entangled state would transform into a mixed quantum state. Bob must perform the proper unitary operations on his qubits to remotely construct the quantum state, therefore the final state $\rho_{out} ^r$  may be written as the density matrix indicated in the following Eq. \eqref{eq:rho_out}.

\begin{equation} 
    \rho_{out}^r = Tr_{i_1i_2...i_{n-1}} \{ \mathcal{U} [\rho_k \otimes  \xi^r(\rho_l)] \mathcal{U}^{\dagger} \}
    \label{eq:rho_out}
\end{equation}
where $Tr_{i_1i_2...i_{n-1}}$ is the partial trace over the qubits $i_1,i_2,...,i_{n-1}$ and $\mathcal{U}$ is the unitary operations, which Bob will apply on his qubit to remotely prepare the quantum state, the operation $\mathcal{U}$ is given by the following Eq. \eqref{eq:U}, $\mathcal{U}$

\begin{eqnarray}
\mathcal{U} &=& \{\mathbb{I}_1 \otimes \mathbb{I}_2 \otimes . . .\mathbb{I}_{n-1} \otimes \sigma_{n} ^{i_1 i_2 ... i_{n-1}} \} \{\ket{\phi}_{12} \bra{\phi}_{12} \otimes \mathbb{I}_3 ... \otimes \mathbb{I}_n \} \{\mathbb{I}_1 \otimes \mathbb{I}_2 ... U_{j_1 j_2} ... \otimes \mathbb{I}_n \} \nonumber \\
&& \{\mathbb{I}_1 \otimes \mathbb{I}_2 ... U_{k_1} ... \otimes \mathbb{I}_n \}
\label{eq:U}
\end{eqnarray}

where $\sigma_{n} ^{i_1 i_2 ... i_{n-1}}$ is the the Bob's recovery unitary operations after Alice has measured her qubit and communicated her result to Bob via a classical channel, $\ket{\phi}_{12} \bra{\phi}_{12}$ is the Bell basis measurement on the first two qubits, $U_{j_1 j_2}$ represents the C-NOT gate from qubit $j_1$ to $j_2$ and $U_{k_1}$ represents the unitary gate on qubit $k_1$. The unitary operations vary according to the collapse states of Alice, Charlie, and David, as given in the table \ref{table:Operation}. By computing the fidelity between the initial two-qubit state $ket{\xi}$ and the density matrix $\rho_{out}^r$, the influence of noise in the entangled channel can now be illustrated. Fidelity reflects the proximity between two quantum states and provides a mathematical formula for quantifying the degree of resemblance between quantum states. The mathematical expression for nose-effected fidelity given by Eq. \eqref{eq:Fid} \cite{liang2019quantum}. 
 \begin{equation}
     \mathcal{F} = \bra{\Psi} \rho_{out}^r \ket{\Psi}
     \label{eq:Fid}
 \end{equation}

Due to the fact that this communication is occurring under the influence of noise, some quantum information may be lost during the remote state preparation. Fidelity is the ideal statistic for measuring how much data is lost. When the fidelity $\mathcal{F}=1$, this indicates the ideal case where the noise hasn't affected the communication, and no information has been lost. Meanwhile, $\mathcal{F}=0$ implies that the noise has very badly affected the communication and the state being communicated has been changed completely to a different state resulting in all the information being lost. Thus, the fidelity ranges between $0$ and $1$, i.e., $0\leq \mathcal{F} \leq 1$. We now discuss the effect of six types of noises (bit-flip, phase-flip, bit-phase-flip, amplitude damping, phase damping and depolarizing noise) in the next section. The entangle channel given in Eq. \eqref{Borras_Eq1} can be factorized in the following manner \eqref{eq-f}

\begin{eqnarray}
    \ket{\Psi} &=& \frac{1}{32}\Big(
    \ket{000}(\ket{0}\ket{\mu^{+}} + \ket{1}\ket{\lambda^{+}})
    +\ket{001}(\ket{0}\ket{\lambda^{-}} - \ket{1}\ket{\mu^{-}})
    +\ket{010}(\ket{0}\ket{\lambda^{+}} - \ket{1}\ket{\mu^{+}}) + \ket{011}(\ket{0}\ket{\mu^{-}} + \ket{1}\ket{\lambda^{-}})
    \nonumber\\ &+& 
    \ket{100}(-\ket{0}\ket{\lambda^{-}} - \ket{1}\ket{\mu^{-}})
    +\ket{101}(\ket{1}\ket{\lambda^{+}}-\ket{0}\ket{\mu^{+}}) 
    +\ket{110}(\ket{0}\ket{\mu^{-}} - \ket{1}\ket{\lambda^{-}})
    +\ket{111}(\ket{0}\ket{\lambda^{+}} + \ket{1}\ket{\mu^{+}}) \Big) \nonumber \\ 
\label{eq-f}    
\end{eqnarray}
where $\ket{\lambda^{\pm}} = \frac{1}{\sqrt{2}} (\ket{011} \pm \ket{100})$ and $\ket{\mu^{\pm}} = \frac{1}{\sqrt{2}} (\ket{000} \pm \ket{111})$ 
\subsection{Bit-flip noisy environment}
The bit-flip noise changes the state of computational qubit $\ket{0}$ to $\ket{1}$ and vice-versa with probability $\eta_b$ and the qubits remain unchanged with the probability $(1-\eta_b)$\cite{fortes2015fighting, oh2002fidelity}. Its operations on a qubit can be described by Kraus operators given by the following matrices in Eq.\eqref{eqn_bit_noise}  

\begin{eqnarray}
 && E_0^b= \sqrt{1-\eta_b} \mathbb{I} =
 \sqrt{1-\eta_b} \begin{pmatrix}
    1 & 0 \\
    0 & 1
  \end{pmatrix}, ~~~~E_1^b= \sqrt{\eta_b}\mathbb{X} = \sqrt{\eta_b}
  \begin{pmatrix}
    0 & 1 \\
    1 & 0
  \end{pmatrix}
\label{eqn_bit_noise}
\end{eqnarray}

Where $ \eta_b \in [0,1]$ represents the probability parameter for the bit-flip error in the quantum system. 
The impact of bit-flip noise on the entangled channel may be analyzed with the help of density matrices of the noisy channel. The affected density matrix under the bit-flip noise  is denoted by $\mathcal{G}^b (\rho)$, given in Eq. \eqref{eq:bit2}.

\begin{eqnarray}
 \mathcal{G}^b (\rho) &=& \frac{1}{32}
 \Big\{(1-\eta_b)^{7} \Big[\Ket{\Psi}\bra{\Psi}\Big]  + (\eta_b)^7
 \Big[\ket{111}(\ket{1}\ket{\mu^{+}} + \ket{0}\ket{\lambda^{+}})
 +\ket{110}(\ket{0}\ket{\mu^{-}} -\ket{1}\ket{\lambda^{-}} )
 +\ket{101}(\ket{1}\ket{\lambda^{+}} - \ket{0}\ket{\mu^{+}}) \nonumber \\&-&
 \ket{100}(\ket{1}\ket{\mu^{-}} + \ket{0}\ket{\lambda^{-}}) + \ket{011}(\ket{1}\ket{\lambda^{-}} - \ket{0}\ket{\mu^{-}}) +\ket{010}(\ket{0}\ket{\lambda^{+}}-\ket{1}\ket{\mu^{+}}) + \ket{001}(\ket{0}\ket{\lambda^{-}} -\ket{1}\ket{\mu^{-}} ) \nonumber \\ &+& 
 \ket{000}(\ket{1}\ket{\lambda^{+}} + \ket{0}\ket{\mu^{+}}) \Big] 
 \times
 \Big[\bra{111}(\bra{1}\bra{\mu^{+}} + \bra{0}\bra{\lambda^{+}}) 
 +\bra{110}(\bra{0}\bra{\mu^{-}} -\bra{1}\bra{\lambda^{-}} )
 +\bra{101}(\bra{1}\bra{\lambda^{+}} - \bra{0}\bra{\mu^{+}}) \nonumber \\ &-& 
 \bra{100}(\bra{1}\bra{\mu^{-}} + \bra{0}\bra{\lambda^{-}}) 
 +\bra{011}(\bra{1}\bra{\lambda^{-}} - \bra{0}\bra{\mu^{-}}) +\bra{010}(\bra{0}\bra{\lambda^{+}}-\bra{1}\bra{\mu^{+}})
 +\bra{001}(\bra{0}\bra{\lambda^{-}} -\bra{1}\bra{\mu^{-}} ) \nonumber \\ &+& 
 \bra{000}(\bra{1}\bra{\lambda^{+}} + \bra{0}\bra{\mu^{+}}) \Big] \Big\} 
     \label{eq:bit2}
\end{eqnarray}


\subsection{Phase-flip noisy environment}
The phase-flip noise affects the phase of the computational qubit. If the system has a phase-flip noise then the computational qubit changes from $\ket{1}$ to $-\ket{1}$, whereas it does not alter the qubit $\ket{0}$. It's Kraus operators \cite{fortes2015fighting, oh2002fidelity} are given by Eq. \eqref{eqn_phase_noise},

\begin{eqnarray}
  && E_0^w= \sqrt{1-\eta_w} \mathbb{I} =
 \sqrt{1-\eta_w} \begin{pmatrix}
    1 & 0 \\
    0 & 1
  \end{pmatrix}~, ~~~~E_1^w= \sqrt{\eta_w}\mathbb{Z} = \sqrt{\eta_w}
  \begin{pmatrix}
    1 & 0 \\
    0 & -1
  \end{pmatrix}
\label{eqn_phase_noise}
\end{eqnarray}
where $\eta_{w} \in [0,1]$ represents the probability parameter for the phase-flip error in the quantum system. The effect of phase-flip noise on the entangled channel can be studied by using the density matrices of the noisy phase-affected channel, which is given by $\mathcal{G}^w (\rho)$
\begin{eqnarray}
\mathcal{G}^w (\rho) &=& \frac{1}{32} \Big\{(1-\eta_w)^{7} \Big[\Ket{\Psi}\bra{\Psi}\Big]  
+ (\eta_w)^7 \Big[\ket{000}(\ket{0}\ket{\mu^{-}} - \ket{1}\ket{\lambda^{-}}) 
- \ket{001} (\ket{0}\ket{\lambda^{+}} +\ket{1}\ket{\mu^{+}} )
- \ket{010}(\ket{0}\ket{\lambda^{-}} + \ket{1} \ket{\mu^{-}}) \nonumber \\
&-& \ket{011}(\ket{0}\ket{\mu^{+}} - \ket{1}\ket{\lambda^{+}}) 
- \ket{100}(\ket{1}\ket{\mu^{+}}-\ket{0}\ket{\lambda^{+}}) 
- \ket{101} (+\ket{0}\ket{\mu^{-}}+\ket{1}\ket{\lambda^{-}}) 
+ \ket{110}(\ket{0}\ket{\mu^{+}} +\ket{1}\ket{\lambda^{+}} ) \nonumber \\ 
&-& \ket{111}(\ket{0}\ket{\lambda^{-}} - \ket{1}\ket{\mu^{-}}) \Big] 
\times
\Big[\bra{000}(\bra{0}\bra{\mu^{-}} - \bra{1}\bra{\lambda^{-}}) 
- \bra{001} (\bra{0}\bra{\lambda^{+}} +\bra{1}\bra{\mu^{+}} )
- \bra{010}(\bra{0}\bra{\lambda^{-}} + \bra{1} \bra{\mu^{-}}) \nonumber \\
&-& \bra{011}(\bra{0}\bra{\mu^{+}} - \bra{1}\bra{\lambda^{+}}) 
- \bra{100}(\bra{1}\bra{\mu^{+}} - \bra{0}\bra{\lambda^{+}} ) 
- \bra{101} (+\bra{0}\bra{\mu^{-}}+\bra{1}\bra{\lambda^{-}}) 
+ \bra{110}(\bra{0}\bra{\mu^{+}} + \bra{1}\bra{\lambda^{+}} ) \nonumber \\ 
&-& \bra{111}(\bra{0}\bra{\lambda^{-}} - \bra{1}\bra{\mu^{-}}) \Big]  \Big\} 
\label{eq:phase}
\end{eqnarray}

\subsection{Bit-phase-flip noisy environment}
The combination of a phase flip and a bit flip is referred to as a bit–phase flip, and it is represented by the Kraus operators given in Eq. \eqref{eqn_bitphase_noise} \cite{fortes2015fighting, oh2002fidelity},

\begin{eqnarray}
 && E^f= \sqrt{1-\eta_f} \mathbb{I} =
  \sqrt{1-\eta_f} \begin{pmatrix}
    1 & 0 \\
    0 & 1
  \end{pmatrix}~, ~~~~ E^f = \sqrt{\eta_f}\mathbb{Y} = \sqrt{\eta_f}
  \begin{pmatrix}
    0 & -i \\
    i & 0
  \end{pmatrix}
\label{eqn_bitphase_noise}
\end{eqnarray}
where $ \eta_{f} \in [0,1]$  denotes the bit-phase-flip noise parameter probability, which defines the likelihood of an error in the quantum state arising owing to a computational qubit. The impact of bit-phase flip noise on the entangled channel can be determined by calculating the noise affected density matrix. The impacted density matrix under bit-phase flip noise is denoted by $\mathcal{G}^f (\rho)$, which is provided by Eq. \eqref{eq:flip}

\begin{eqnarray}
\mathcal{G}^f (\rho) &=& \frac{1}{32} \Big\{(1-\eta_f)^{7} \Big[\Ket{\Psi}\bra{\Psi}\Big] -(\eta_f)^7 \Big[\ket{111}(-\ket{1}\ket{\mu^{-}}+\ket{0}\ket{\lambda^{-}})
-\ket{110}(\ket{0}\ket{\mu^{+}}+\ket{1}\ket{\lambda^{+}})
+\ket{101}(\ket{1}\ket{\lambda^{-}} + \ket{0}\ket{\mu^{-}}) \nonumber \\ &+& 
\ket{100}(\ket{1}\ket{\mu^{+}} - \ket{0}\ket{\lambda^{+}}) - 
\ket{011}(\ket{0}\ket{\mu^{+}} - \ket{1}\ket{\lambda^{+}}) 
+\ket{010}(\ket{0}\ket{\lambda^{-}}+\ket{1}\ket{\mu^{-}}) 
+\ket{001}(\ket{0}\ket{\lambda^{+}} + \ket{1}\ket{\mu^{+}}) \nonumber \\ &-& 
\ket{000}(\ket{0}\ket{\lambda^{-}} + \ket{0}\ket{\mu^{-}}) \Big] 
\times
\Big[\bra{111}(-\bra{1}\bra{\mu^{-}}+ \bra{0}\bra{\lambda^{-}}) 
-\bra{110}(\bra{0}\bra{\mu^{+}}+\bra{1}\bra{\lambda^{+}})
+\bra{101}(\bra{1}\bra{\lambda^{-}} + \bra{0}\bra{\mu^{-}}) \nonumber \\ &+&
\bra{100}(\bra{1}\bra{\mu^{+}} - \bra{0}\bra{\lambda^{+}}) 
-\bra{011}(\bra{0}\bra{\mu^{+}} - \bra{1}\bra{\lambda^{+}}) 
+\bra{010}(\bra{0}\bra{\lambda^{-}}+\bra{1}\bra{\mu^{-}}) 
+\bra{001}(\bra{0}\bra{\lambda^{+}} + \bra{1}\bra{\mu^{+}}) \nonumber \\ &-& 
\bra{000}(\bra{0}\bra{\lambda^{-}} + \bra{0}\bra{\mu^{-}}) \Big] \Big\} 
\label{eq:flip}
\end{eqnarray}

\subsection{Effect of amplitude damping (AD) noise}
The amplitude damping plays a vital role since it is responsible for characterising the energy loss of a system, which is the effect that occurs the most frequently in open systems. The idea of amplitude damping is crucial to the modelling of energy dissipation in a variety of quantum systems, and the matrices that follow supply the Kraus operators for this process \cite{fortes2015fighting, oh2002fidelity}.
\begin{eqnarray}
  E_0^a=
  \begin{pmatrix}
    1 & 0 \\
    0 & \sqrt{1-\eta_{a}}
  \end{pmatrix}~, \quad
  E_1^a=
  \begin{pmatrix}
    0 & \sqrt{\eta_{a}} \\
    0 & 0
  \end{pmatrix}
\label{eqn_AD_noise}
\end{eqnarray}
Where $\eta_{a} \in [0,1]$ signifies the decoherence rate of amplitude damping, which specifies the likelihood of quantum state inaccuracy associated with computational qubits. The influence of amplitude damping on the entangled channel may be detected by analysing the channel's noise-affected density matrix. This matrix is represented by $\mathcal{G}^a$, given in Eq. \eqref{eq:AP}

\begin{eqnarray}
\mathcal{G}^a (\rho) &=& \frac{1}{32} \Big\{ \Big[\ket{000}(\ket{0}\ket{\Xi^{+}}+\sqrt{1-\eta_a}\ket{1} \ket{\Omega^{+}}) +\sqrt{1-\eta_a}\ket{001}(\ket{0}\ket{\Omega^{-}}-\sqrt{1-\eta_a}\ket{1}\ket{\Xi^{-}}) \nonumber \nonumber \\
&+& \sqrt{1-\eta_a}\ket{010}(\ket{0}\ket{\Omega^{+}} - \sqrt{1-\eta_a}\ket{1}\ket{\Xi^{+}}) + 
(1-\eta_a) \ket{011}(\ket{0}\ket{\Xi^{-}} + \sqrt{1-\eta_a}\ket{1}\ket{\Omega^{-}}) \nonumber \\ 
&-& \sqrt{1-\eta_a}\ket{100}(\ket{0}\ket{\Omega^{-}} + \sqrt{1-\eta_a} \ket{1}\ket{\Xi^{-}}) 
+(1-\eta_a)\ket{101}(\sqrt{1-\eta_a}\ket{1}\ket{\Omega^{+}}-\ket{0}\ket{\Xi^{+}}) \nonumber \\
&+& (1-\eta_a)\ket{110}(\ket{0}\ket{\Xi^{-}} - \sqrt{1-\eta_a}\ket{1}\ket{\Omega^{-}}) + 
\sqrt{(1-\eta_a)^3}\ket{111}(\ket{0}\ket{\Omega^{+}} + \sqrt{1-\eta_a}\ket{1}\ket{\Xi^{+}}) \Big] \nonumber \\ 
&\times& \Big[\bra{000}(\bra{0}\bra{\Xi^{+}}+\sqrt{1-\eta_a}\bra{1} \bra{\Omega^{+}}) +\sqrt{1-\eta_a}\bra{001}(\bra{0}\bra{\Omega^{-}}-\sqrt{1-\eta_a}\bra{1}\bra{\Xi^{-}}) \nonumber \\  
&+& \sqrt{1-\eta_a}\bra{010}(\bra{0}\bra{\Omega^{+}} - \sqrt{1-\eta_a}\bra{1}\bra{\Xi^{+}}) \nonumber + 
(1-\eta_a) \bra{011}(\bra{0}\bra{\Xi^{-}} + \sqrt{1-\eta_a}\bra{1}\bra{\Omega^{-}}) \nonumber \\
&-& \sqrt{1-\eta_a}\bra{100}(\bra{0}\bra{\Omega^{-}} + \sqrt{1-\eta_a} \bra{1}\bra{\Xi^{-}}) 
+(1-\eta_a)\bra{101}(\sqrt{1-\eta_a}\bra{1}\bra{\Omega^{+}}-\bra{0}\bra{\Xi^{+}}) \nonumber \\
&+& (1-\eta_a)\bra{110}(\bra{0}\bra{\Xi^{-}} - \sqrt{1-\eta_a}\bra{1}\bra{\Omega^{-}})  +
\sqrt{(1-\eta_a)^3}\bra{111}(\bra{0}\bra{\Omega^{+}} + \sqrt{1-\eta_a}\bra{1}\bra{\Xi^{+}}) \Big] \nonumber \\ &+& (\eta_a)^7 \Ket{1111111}\bra{1111111} \Big\} 
\label{eq:AP}
\end{eqnarray}

where the quantum states $\ket{\Omega^{\pm}} = \dfrac{\sqrt{1-\eta_a}\big(\sqrt{1-\eta_a}\Ket{011} \pm \Ket{100} \big)}{\sqrt{2}}$ and $\ket{\Xi^{\pm}} = \dfrac{\big(\Ket{000} \pm \sqrt{(1-\eta_a)^3}\Ket{111} \big)}{\sqrt{2}}$

\subsection{Phase damping noisy environment}
The phenomenon of phase-damping results in the loss of information regarding the relative phases of a quantum state. During phase damping, the fundamental quantum system becomes entangled with the surrounding environment \cite{mcmahon2007quantum}. The Kraus operators \{$E_{0}^p, E_{1}^p, E_{2}^p$\} for phase-damping noise are given in Eq. \eqref{eqn_PD_noise} \cite{fortes2015fighting, oh2002fidelity}.
\begin{eqnarray}
  && E_0^p = \sqrt{1-\eta_{p}}
  \begin{pmatrix}
    1 & 0 \\
    0 & 1
  \end{pmatrix}~, \quad
  E_1^p=
  \begin{pmatrix}
    \sqrt{\eta_{p}} & 0 \\
    0 & 0
  \end{pmatrix}~, \quad ~~~ E_2^p=
  \begin{pmatrix}
      0 & 0 \\
      0 & \sqrt{\eta_{p}}
  \end{pmatrix}
\label{eqn_PD_noise}
\end{eqnarray}
where $\eta_{p} \in [0,1]$ denotes the phase-damping decoherence rate, which specifies the likelihood of an error occurring in the quantum state associated with the computational qubit. After the noise has been introduced into the channel, the affected density matrix can be used to figure out how phase damping affected the entangled channels. The affected density matrix under the phase damping noise is denoted by $\mathcal{G}^p$.

\begin{eqnarray}
 \mathcal{G}^p (\rho) &=& \frac{1}{32}
 \Big\{(1-\eta_b)^{7} \Big[\Ket{\Psi}\bra{\Psi}\Big]  + (\eta_p)^7
 \Big[\ket{0000000}\bra{0000000} + \Ket{1111111}\bra{1111111}\Big] \Big\} 
     \label{eq:PD}
\end{eqnarray}


\subsection{Depolarizing Noisy Environment}
When exposed to a depolarizing noisy environment, the quantum state's qubits are depolarized with a probability of $\eta_d$, and the qubits are left with an invariant probability of $(1-\eta_d)$.  The following matrices give the Kraus operators for depolarizing noise \cite{fortes2015fighting, oh2002fidelity}.

\begin{eqnarray}
 && E_0^d = \sqrt{1-\eta_{d}}
  \begin{pmatrix}
    1 & 0 \\
    0 & 1
  \end{pmatrix}~, \quad
  E_1^d = \sqrt{\frac{\eta_{d}}{3}}
  \begin{pmatrix}
    0 & 1 \\
    1 & 0
  \end{pmatrix}~, ~~~ E_2^d = \sqrt{\frac{\eta_{d}}{3}}
  \begin{pmatrix}
      0 & -i \\
      i & 0
  \end{pmatrix}~, \quad
  E_3^d = \sqrt{\frac{\eta_{d}}{3}}
  \begin{pmatrix}
      1 & 0 \\
      0 & -1
  \end{pmatrix}
\label{eqn_DN_noise}
\end{eqnarray}

After introducing noise into a channel, the impact of depolarizing noise on the entangled channel may be determined by analysing the affected density matrix, the noise affected density matrix under the depolarizing noise is denoted by $\mathcal{G}^D$
\begin{eqnarray}
\mathcal{G}^d (\rho) &=& \frac{1}{32} \Big\{(1-\eta_d)^{7} 
\Big[\Ket{\Psi}\bra{\Psi}\Big] + (\eta_b)^7
\Big[\ket{111}(\ket{1}\ket{\mu^{+}} + \ket{0}\ket{\lambda^{+}})
+\ket{110}(\ket{0}\ket{\mu^{-}} -\ket{1}\ket{\lambda^{-}} )
+\ket{101}(\ket{1}\ket{\lambda^{+}} - \ket{0}\ket{\mu^{+}}) \nonumber \\&-&
\ket{100}(\ket{1}\ket{\mu^{-}} + \ket{0}\ket{\lambda^{-}}) 
+ \ket{011}(\ket{1}\ket{\lambda^{-}} - \ket{0}\ket{\mu^{-}}) +\ket{010}(\ket{0}\ket{\lambda^{+}}-\ket{1}\ket{\mu^{+}}) 
+ \ket{001}(\ket{0}\ket{\lambda^{-}} -\ket{1}\ket{\mu^{-}} ) \nonumber \\ &+& 
\ket{000}(\ket{1}\ket{\lambda^{+}} + \ket{0}\ket{\mu^{+}}) \Big]
\times
\Big[\bra{111}(\bra{1}\bra{\mu^{+}} + \bra{0}\bra{\lambda^{+}}) 
+\bra{110}(\bra{0}\bra{\mu^{-}} -\bra{1}\bra{\lambda^{-}} )
+\bra{101}(\bra{1}\bra{\lambda^{+}} - \bra{0}\bra{\mu^{+}}) \nonumber \\ &-& 
\bra{100}(\bra{1}\bra{\mu^{-}} + \bra{0}\bra{\lambda^{-}}) 
+\bra{011}(\bra{1}\bra{\lambda^{-}} - \bra{0}\bra{\mu^{-}}) +\bra{010}(\bra{0}\bra{\lambda^{+}}-\bra{1}\bra{\mu^{+}})
+\bra{001}(\bra{0}\bra{\lambda^{-}} -\bra{1}\bra{\mu^{-}} ) \nonumber \\ &+& 
\bra{000}(\bra{1}\bra{\lambda^{+}} + \bra{0}\bra{\mu^{+}}) \Big]-(\eta_f)^7 \Big[\ket{111}(-\ket{1}\ket{\mu^{-}}+\ket{0}\ket{\lambda^{-}})
-\ket{110}(\ket{0}\ket{\mu^{+}}+\ket{1}\ket{\lambda^{+}})
+\ket{101}(\ket{1}\ket{\lambda^{-}} + \ket{0}\ket{\mu^{-}}) \nonumber \\ &+& 
\ket{100}(\ket{1}\ket{\mu^{+}} - \ket{0}\ket{\lambda^{+}}) - 
\ket{011}(\ket{0}\ket{\mu^{+}} - \ket{1}\ket{\lambda^{+}}) 
+\ket{010}(\ket{0}\ket{\lambda^{-}}+\ket{1}\ket{\mu^{-}}) 
+\ket{001}(\ket{0}\ket{\lambda^{+}} + \ket{1}\ket{\mu^{+}}) \nonumber \\ &-& 
\ket{000}(\ket{0}\ket{\lambda^{-}} + \ket{0}\ket{\mu^{-}}) \Big] 
\times
\Big[\bra{111}(-\bra{1}\bra{\mu^{-}}+ \bra{0}\bra{\lambda^{-}}) 
-\bra{110}(\bra{0}\bra{\mu^{+}}+\bra{1}\bra{\lambda^{+}})
+\bra{101}(\bra{1}\bra{\lambda^{-}} + \bra{0}\bra{\mu^{-}}) \nonumber \\ &+&
\bra{100}(\bra{1}\bra{\mu^{+}} - \bra{0}\bra{\lambda^{+}}) 
-\bra{011}(\bra{0}\bra{\mu^{+}} - \bra{1}\bra{\lambda^{+}}) 
+\bra{010}(\bra{0}\bra{\lambda^{-}}+\bra{1}\bra{\mu^{-}}) 
+\bra{001}(\bra{0}\bra{\lambda^{+}} + \bra{1}\bra{\mu^{+}}) \nonumber \\ &-& 
\bra{000}(\bra{0}\bra{\lambda^{-}} + \bra{0}\bra{\mu^{-}}) \Big] 
+ (\eta_w)^7 \Big[\ket{000}(\ket{0}\ket{\mu^{-}} - \ket{1}\ket{\lambda^{-}}) 
- \ket{001} (\ket{0}\ket{\lambda^{+}} +\ket{1}\ket{\mu^{+}} )
- \ket{010}(\ket{0}\ket{\lambda^{-}} + \ket{1} \ket{\mu^{-}}) \nonumber \\&-&
\ket{011}(\ket{0}\ket{\mu^{+}} - \ket{1}\ket{\lambda^{+}}) 
- \ket{100}(\ket{1}\ket{\mu^{+}}-\ket{0}\ket{\lambda^{+}}) - \ket{101} (+\ket{0}\ket{\mu^{-}}+\ket{1}\ket{\lambda^{-}}) 
+ \ket{110}(\ket{0}\ket{\mu^{+}} + \ket{1}\ket{\lambda^{+}} ) \nonumber \\ 
&-& \ket{111}(\ket{0}\ket{\lambda^{-}} - \ket{1}\ket{\mu^{-}}) \Big] 
\times 
\Big[\bra{000}(\bra{0}\bra{\mu^{-}} - \bra{1}\bra{\lambda^{-}}) 
- \bra{001} (\bra{0}\bra{\lambda^{+}} +\bra{1}\bra{\mu^{+}} )
- \bra{010}(\bra{0}\bra{\lambda^{-}} + \bra{1} \bra{\mu^{-}}) \nonumber \\ &-&
\bra{011}(\bra{0}\bra{\mu^{+}} - \bra{1}\bra{\lambda^{+}}) 
- \bra{100}(\bra{1}\bra{\mu^{+}} - \bra{0}\bra{\lambda^{+}} ) 
- \bra{101} (+\bra{0}\bra{\mu^{-}}+\bra{1}\bra{\lambda^{-}}) 
+ \bra{110}(\bra{0}\bra{\mu^{+}} + \bra{1}\bra{\lambda^{+}} ) \nonumber \\ 
&-& \bra{111}(\bra{0}\bra{\lambda^{-}} - \bra{1}\bra{\mu^{-}}) \Big]\Big\} 
\label{eq:Depo}
\end{eqnarray}

\begin{figure}[H]
    \centering
    \includegraphics[width=0.6\textwidth]{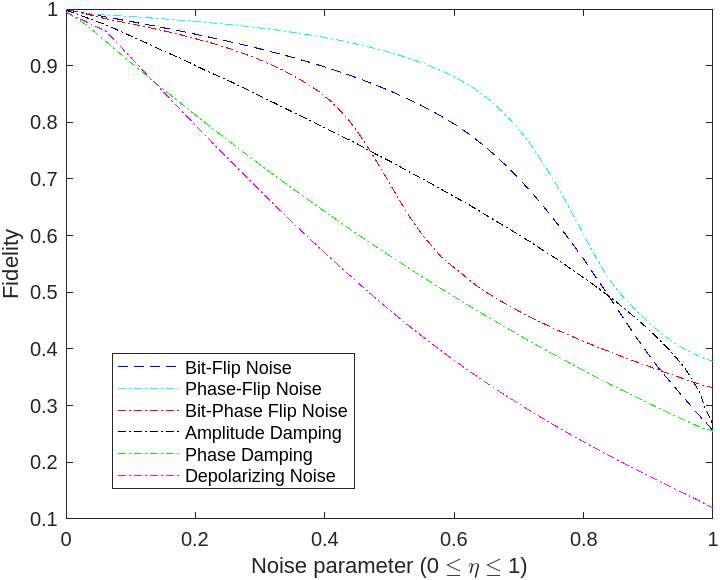}
    \caption{The plot of fidelity against the noise parameter $\eta$ for all the six kinds of noise models.}
    \label{fig:RSPFig}
\end{figure}

\section{Security analysis}\label{Section-4}
Quantum communication protocols, in comparison to their classical counterparts, often have the characteristic of a superior level of security. Our protocols are protected not just from attacks coming from the outside, but also from those coming from inside the system from the dishonest participant. Here, we provide two different types of security analysis about the process of remotely preparing a state. The first of these is an outside attack from an eavesdropper, attempting to learn the state that is being remotely prepared. And the second threat is an attempt from a non-authorized participant who is interested in finding out the quantum state that is being prepared remotely.

\subsection{Outside attack}
Before the remote state preparation scheme is put into action, the seven-quit entangled channel is to be created from the six-qubit Borras \emph{et al.} state and then distributed among the participants. Without loss of generality, suppose Alice prepares the state $\Ket{\Psi}_{1234567}$ and send the qubits $(2,3)$ to Bob, qubits $(4,6)$ to Charlie and $(5,7)$ to David. Alice includes a predetermined number of decoy-state particles in the transmission which are randomly distributed in one of the following four quantum states $\{ \Ket{0}, \Ket{1}, \Ket{+}, \Ket{-} \}$.
After it has been determined that all three participants have received the particles, Alice will then proceed to make the statement about the placements of the decoy particles as well as the measurement basis. Then, participants Bob, Charlie, and David measure their qubits in accordance with the provided basis and declare their results. Next, Alice makes a comparison between the results of the measurement and the initial states of the decoy particles. The fact that any eavesdropping leaves a trace in the outcomes of the decoy sampling photons \cite{wang2008efficient}, enables the security checking process to detect multiple types of attacks coming from an outside attacker Eve. These attacks include several attacks mentioned in \cite{chen2012controlled} named as an intercept-resend attack, a measurement-resend attack, an entanglement-measure attack, and a denial-of-service attack. During the security checking, the impact of these attacks will be detected with a probability larger than zero. This verification approach is based on the notion of the BB84 QKD protocol \cite{bennett2020quantum}, which has been shown to be completely safe by a number of different researchers  \cite{shor2000simple}. 
Moreover, the particles used to construct the quantum channel do not transmit any concealed information. As a result, if an eavesdropper is present, she is not only identifiable but also incapable of gathering any relevant information during the security screening process. After the completion of the security checks, the entangled channel that is sufficiently safe will be distributed among the participants. If there is evidence of eavesdropping, the participants will abandon this procedure and begin over. After three different parties have validated the safety of the quantum channel, an eavesdropper from the outside can no longer attack the protocol since no qubits are being sent at this point. During the implementation of the protocol, only classical information is sent, which has no relevance to the secrets. Therefore, our remote state preparation protocol is robust against an attack from an outsider.

\subsection{Inside attack}
It is possible for a participant to carry out his attack by entangling an auxiliary particle with his own particle. Let's suppose that Alice is the participant interested in finding out about Bob's state, being remotely prepared in an unethical way. She can prepare the auxiliary state $\Ket{\varepsilon}$ and entangle it using the local unitary operation $\mathcal{\hat{U}}$, which is defined as follows:

\begin{eqnarray}
 \mathcal{\hat{U}} (\Ket{0}_1\Ket{\varepsilon}_E) &=& \Ket{0}_1 \Ket{\varepsilon_{00}} + \Ket{1}_1 \Ket{\varepsilon_{01}} \nonumber \\ 
 \mathcal{\hat{U}} (\Ket{1}_1\Ket{\varepsilon}_E) &=& \Ket{0}_1 \Ket{\varepsilon_{10}} + \Ket{1}_1 \Ket{\varepsilon_{11}}
 \label{eq:EntQ}
\end{eqnarray}

Where $\braket{\varepsilon_{00}|\varepsilon_{00}} + \braket{\varepsilon_{01}|\varepsilon_{01}} = 1 $ and
$\braket{\varepsilon_{10}|\varepsilon_{10}} + \braket{\varepsilon_{11}|\varepsilon_{11}} = 1$.
Suppose Charlie and David simultaneously measure their qubits and broadcast their results to Bob through a classical channel. Without compromising generality, assume that the Charlie and David's measurement results are $\Ket{00}_{C_1C_2}\Ket{00}_{D_1D_2}$. From table \ref{table:Operation}, Alice's measurement result is  $\ket{\Upsilon_1}_A$ or $\ket{\Upsilon_2}_A$. After completing the unitary operations \eqref{eq:EntQ} on her state, Alice entangled an auxiliary qubit on it, and the resulting state becomes

\begin{eqnarray}
 \ket{\Psi}_{AEB_1B_2} &=& \frac{1}{4} \Big( \mathcal{\hat{U}} \Ket{\Upsilon_1}_A \Ket{\varepsilon}_E\Big) \otimes \Big[ \frac{1}{\sqrt{2}}(\alpha \Ket{00} + \beta \Ket{10} + \alpha \Ket{11} - \beta \Ket{01}) \Big] \nonumber \\
 &+& \frac{1}{4} \Big( \mathcal{\hat{U}} \Ket{\Upsilon_2}_A \Ket{\varepsilon}_E \Big) \otimes \Big[ \frac{1}{\sqrt{2}}(\alpha \Ket{10} - \beta \Ket{00} - \alpha \Ket{01} - \beta \Ket{11}) \Big] \nonumber \\
 &=& \frac{1}{4} \Big[ \alpha (\Ket{0}\Ket{\varepsilon_{00}} + \Ket{1}\Ket{\varepsilon_{01}}) + \beta (\Ket{0} \Ket{\varepsilon_{10}} +\Ket{1} \ket{\varepsilon_{11}}) \Big] \Big[ \alpha \Ket{\psi^+} - \beta \Ket{\phi^-} \Big]
 \nonumber \\ &+& 
 \frac{1}{4} \Big[ \alpha (\Ket{0}\Ket{\varepsilon_{10}} + \Ket{1}\Ket{\varepsilon_{11}}) - \beta (\Ket{0} \Ket{\varepsilon_{00}} +\Ket{1} \ket{\varepsilon_{01}}) \Big] \Big[-\beta \Ket{\psi^+} - \alpha\Ket{\phi^-} \Big]
 \label{eq:EntQ2}
\end{eqnarray}

At this point, Alice must follow the rules for preparing the remote state and tell Bob the result of her measurement. Bob will then do the unitary operations on his qubit. Now, we will concentrate on the quantum system of particles $AE$, which represents a portion of the entire system.

\begin{eqnarray}
 \rho_{AE} &=& tr_{B_1B_2} \big( \rho_{AEB_1B_2} \big) \nonumber \\
 &=& \dfrac{(\alpha^4 + \beta^4)}{16} \Big[ (\Ket{0}\Ket{\varepsilon_{00}} + \Ket{1}\Ket{\varepsilon_{01}}) (\bra{0}\bra{\varepsilon_{00}} + \bra{1}\bra{\varepsilon_{01}})  + (\Ket{0}\Ket{\varepsilon_{10}} + \Ket{1}\Ket{\varepsilon_{11}}) (\bra{0}\bra{\varepsilon_{10}} + \bra{1}\bra{\varepsilon_{11}}) \Big] \nonumber \\
 &=& \dfrac{(\alpha^4 + \beta^4)}{16} \Big( (\Ket{0\varepsilon_{00}} + \Ket{1\varepsilon_{01}}) (\bra{0\varepsilon_{00}} + \bra{1\varepsilon_{01}}) +  (\Ket{0\varepsilon_{10}} + \Ket{1\varepsilon_{11}}) (\bra{0\varepsilon_{10}} + \bra{1\varepsilon_{11}}) \Big)
 \label{eq:EntQ3}
\end{eqnarray}

Now, we will check the purity of the quantum state by evaluating the trace of the square of the density matrix, given by 

\begin{eqnarray}
 tr(\rho^2_{AE}) &=& \Big(\dfrac{\alpha^4 + \beta^4}{16}\Big)^2 + \Big(\dfrac{\alpha^4 + \beta^4}{16}\Big)^2 + 2\Big( \dfrac{\alpha^4 + \beta^4}{16} \Big) \Big( \dfrac{\alpha^4 + \beta^4}{16} \Big) |\braket{\varepsilon_{00}|\varepsilon_{10}} + \braket{\varepsilon_{01}|\varepsilon_{11}}|^2 \nonumber \\
 &\leq& 2 \Big(\dfrac{\alpha^4 + \beta^4}{16}\Big)^2 \Big[1+(|\varepsilon_{00}|\varepsilon_{10}| + |\varepsilon_{10}|\varepsilon_{11}|)^2] \nonumber \\
 &\leq& 2 \Big(\dfrac{\alpha^4 + \beta^4}{16}\Big)^2 \Big[1+(\sqrt{\braket{\varepsilon_{00}|\varepsilon_{00}} \braket{\varepsilon_{10}|\varepsilon_{10}}} + \sqrt{\braket{\varepsilon_{01}|\varepsilon_{01}} \braket{\varepsilon_{11}|\varepsilon_{11}}})^2 \Big] \nonumber \\
 &\leq& 2 \Big(\dfrac{\alpha^4 + \beta^4}{16}\Big)^2 \Big[1+\Big( \dfrac{\braket{\varepsilon_{00}|\varepsilon_{00}} +\braket{\varepsilon_{10}|\varepsilon_{10}}}{2} + \dfrac{\braket{\varepsilon_{01}|\varepsilon_{01}} 
 +\braket{\varepsilon_{11}|\varepsilon_{11}}}{2} \Big)^2 \Big]
 \label{eq:EntQ4}
\end{eqnarray}


Simplifying the expression using the Cauchy–Schwarz inequality and the triangular inequality, we get
\begin{eqnarray}
 tr(\rho^2_{AE}) < 1 
 \label{eq:EntQ6}
\end{eqnarray}
If this is the case, the quantum state of the particle $AE$ will be in a mixed state, and Alice will not be able to extract any information from the system. That is to say, the inner attack initiated by Alice cannot be considered valid and Alice will not be able to steal any information from the attack. 



\section{Discussion and Conclusion }\label{Section-5}
 In this study, the deterministic remote state preparation is studied via the seven-qubit entangled channel, created from the maximally entangled Borras \emph{et al.} state under the effect of six different kinds of noises. First, the seven-qubit entangled channel was constructed from the highly entangled Borras state by taking an ancillary qubit and then performing a C-NOT operation from the terminal qubit of the Borras state to the ancillary qubit. This was done in order to establish the highly entangled Borras state. It generates an entangled channel in order to facilitate the remote state preparation of a two-qubit quantum state. The qubits that make up the entangled channels are split up among the four participants. Alice is the one who is sending the information, while Bob is the one who is participating and preparing the state at his end remotely. After Alice and the other two players, Charlie and David, have sent their measurement results to Bob using a classical channel, Bob then performs the relevant operation to his qubit in order to remotely prepare the state. The operations in the different scenarios are presented in table \ref{table:Operation}.
 In addition, the impacts of six distinct types of noise are investigated during the course of this study. The noise may be modelled using the Kraus operators by applying them to the entangled channel. We have evaluated the density matrices for all the noise channels and calculated the fidelity of the remotely prepared state with respect to the noise-affected channel.
 For various noise channels, it is possible to discern how the system loses information due to decoherence when it interacts with its environment.  A fidelity metric has been used to calculate the amount of information that is lost due to each type of noise, for which variation of fidelity with respect to the noise parameter $\eta$ is plotted in figure \ref{fig:RSPFig}. The plot shows that as the noise parameter increases in the range $[0,1]$, the fidelity shows different trends in different types of noise. The graph reveals that depolarizing noise has the most negative effect on the channel, hence reducing its fidelity and causing information loss. In addition, the phase flip noise loses the least amount of information compared to all others. Moreover, the security analysis is conducted for a protocol intruder. In the event of an external attack, an eavesdropper may attempt to attack the protocol, tamper with the process, and determine the prepared quantum state. However, it appears that our protocol is secure against these types of threats, as outsiders cannot steal information. Again, in the event of an inside attack, a dishonest player may attempt to steal information by entangling an auxiliary qubit $\Ket{\varepsilon}$ with his qubit in order to determine the state being prepared. Attempting to do so, however, will result in the quantum state transforming into a mixed state, resulting in the loss of information.
\bibliography{Arxiv} 
\bibliographystyle{ieeetr}
\end{document}